\documentclass[english,two column, showpacs]{revtex4}
\usepackage[T1]{fontenc}
\usepackage[latin1]{inputenc}
\usepackage{graphicx}
\usepackage{amsmath}
\usepackage{amssymb}
\usepackage{amsfonts}
\def\a{\alpha}

\def\e{\varepsilon}
\def\d{\delta}

\def\r{\rho}

\def\D{\Delta}

\def\be{\begin{equation}}
\def\ee{\end{equation}}
\def\bea{\begin{eqnarray}}
\def\eea{\end{eqnarray}}
\def\nn{\nonumber}
\def\lb{\label}

\begin{document}

\title{Modulated electronic configurations in selectively doped multilayered nanostructures}
\author{V. M. Loktev$^1$, Yu. G. Pogorelov$^2$}
\affiliation{$^1$N.N. Bogolyubov Institute for Theoretical Physics, Metrologichna 14b, Kiev
03134, Ukraine,\\ $^2$IFIMUP, Universidade do Porto, R. Campo Alegre, 687, Porto 4169-007}

\begin{abstract}
A simple theoretical model is proposed to describe the recent experimental results on
formation of induced superconducting state and anomalous tunneling characteristics in
selectively doped multilayered nanostructures based on La$_2$CuO$_4$ perovskite. In particular,
it is shown that the structure composed from the nominally non-superconducting (undoped and
overdoped) layers turns to be superconducting with superconductivity confined to narrow
regions near the interfaces, in agreement with the experimental observations.
\end{abstract}

\pacs{73.40.Gk, 73.50.-h, 73.61.-r}
\maketitle\

Recent experiments by Bozovic \emph{et al} \cite{Boz1} provided a new intriguing insight on
electronic properties of nanostructured perovskite systems. Using thorough epitaxy techniques
available in Brookhaven National Laboratory \cite{Boz2,Boz3,Goz}, they were able to selectively
introduce a well controlled level (including zero) of Sr dopants into each particular
La$_2$CuO$_4$ layer (along the \emph{c}-axis) and then observed unusual electronic characteristics
of the composite structures. For instance, a stack of 15 alternating $\left({\rm La}_{2-x}{\rm
Sr}_x{\rm CuO}_4\right)_4\left({\rm La}_2{\rm CuO}_4\right)_2$ blocks with $x = 0.45$, that is
alternating overdoped \cite{kast} and undoped (both separately non-superconducting) layers,
revealed superconductivity with the critical temperature $T_c = 30$ K \cite{Boz1}. The authors
interpreted this behavior as an evidence for carriers delocalization beyond the nominally doped
region of the multilayered system. Below we propose a very simple theoretical model permitting a
qualitative and semi-quantitave explanation of such delocalization effect.

The euristic basis for the model is the assumption that the collective electronic states in
the multilayered system are superpositions of almost uncoupled (because of a very slow
\emph{c}-axis hopping $t_c$) planar states in each \emph{j}th ${\rm La}_2{\rm CuO}_4$ layer,
formed by the fast \emph{ab}-hopping $t_{ab} \gg t_c$ in the energy band of width $W = 8t_{ab}$
around the relevant atomic level and shifted by a certain local electric potential $\varphi_j$.
The latter is related to the local charge densities $\r_j = e\left(p_j - x_j\right)$ by mobile
holes with density $p_j$ and ionized dopants with density $x_j$ (where $e$ is the elementary
charge), accordingly to the discrete version of the common Poisson equation:
\be
 \varphi_{j+1} + \varphi_{j-1} - 2\varphi_j = -\frac{4\pi c}{\e_{\rm eff}a^2}\r_j,
  \lb{eq1}
   \ee
where $a$ and $c$ are the in-plane and \emph{c}-axis lattice parameters and $\e_{\rm eff}$
is the (static) dielectric constant that effectively reduces the Coulomb field in the
\emph{c}-direction. This equation would be exact for potentials in a stack of mathematical
planes, with uniform charge densities $\r_j$ and separation $c$, and should model real
${\rm La}_{2-x}{\rm Sr}_x{\rm CuO}_4$ layers where $p_j$ delocalized holes and $x_j$ localized
dopants are distributed in different atomic planes within the period $c$ of \emph{j}th layer.
The adopted form of purely dielectric screening is justified in neglect of \emph{c}-hopping
processes, accordingly to their above mentioned weakness. We note that the charge densities
$\r_j$ naturally vanish both in uniformly doped ($p_j = x_j$) and undoped ($p_j = x_j = 0$)
systems.

\begin{figure}
\center\includegraphics[width=8cm]{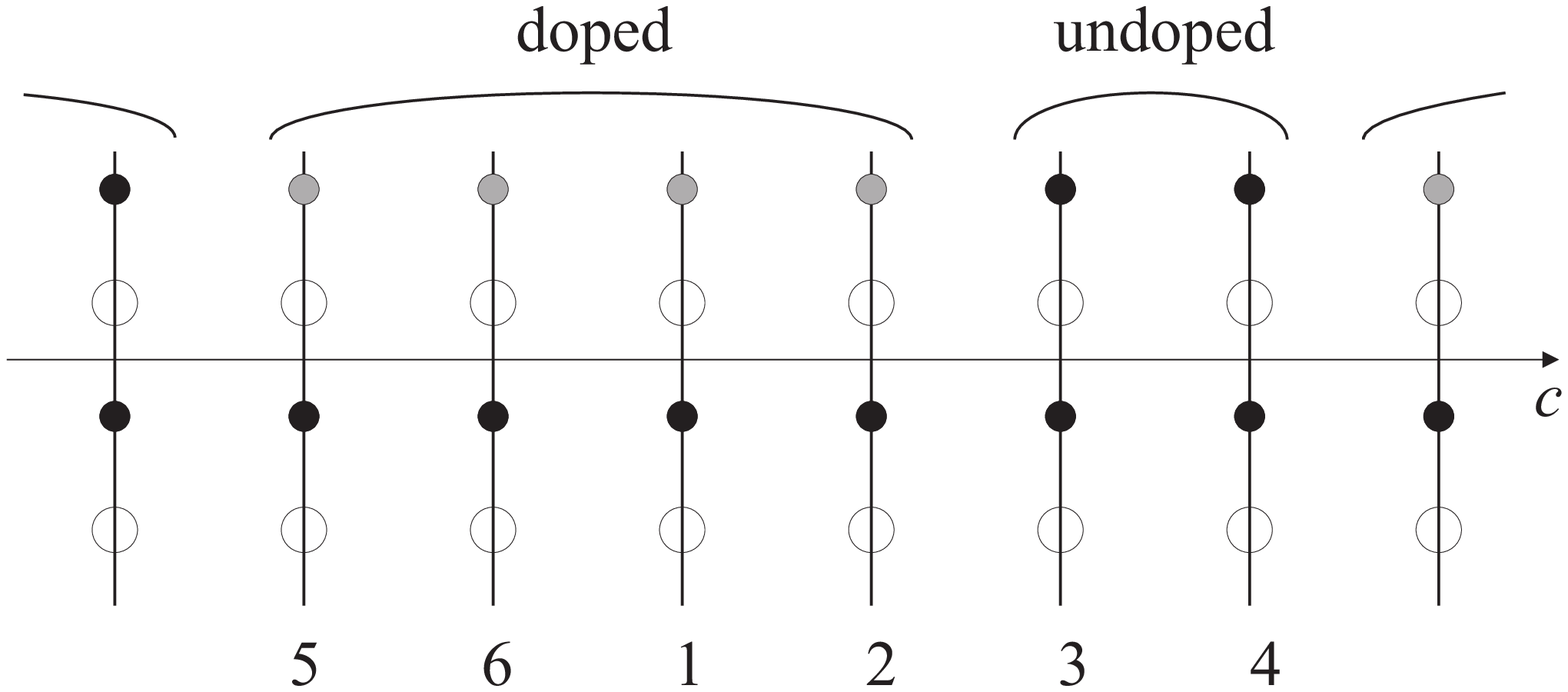}\\
\caption{Schematic of nanostructured system with periodically introduced dopants (light grey
circles) into consecutive layers of ${\rm La}_2{\rm CuO}_4$ along the \emph{c}-axis. There are
only 3 independent values of electronic density over 6 layers in a period.}
  \label{fig1}
\end{figure}

Otherwise, the hole carrier density $p_j$ is defined by the respective density of states
(DOS) $g_j(\e)$:
\be
 p_j = 2\int_{\e_{\rm F}}^{W/2 - e\varphi_j}g_j(\e)d\e,
  \lb{eq2}
   \ee
including the spin factor 2. Thus the role of \emph{c}-hopping in this model is reduced to
establishing the common Fermi level $\e_{\rm F}$ for all the layers. Using the simplest
approximation of rectangular DOS: $g_j(\e) = 1/W$ within the bandwidth $W$, we arrive at the
linear relation between $p_j$ and $\varphi_j$:
\be
 e\varphi_j = \frac{1-p_j} 2 W  - \e_{\rm F}.
  \lb{eq3}
   \ee
Then inserting Eq. \ref{eq3} into Eq. \ref{eq1} leads to a non-uniform linear system for
the densities $p_j$:
\be
 p_{j+1} + p_{j-1} - (2 + \a)p_j = -\a x_j,
  \lb{eq4}
   \ee
where the dimensionless value:
\be
 \a = \frac{8\pi c e^2}{W\e_{\rm eff}a^2}
  \lb{eq5}
   \ee
is a single material parameter of the model, describing the localization degree of charge
density fluctuations in the nanostructured system (less delocalization for bigger $\a$).
The advantage of Eq. \ref{eq4} against an analogous system for potentials $\varphi_j$ is
in eliminating the Fermi level (doping dependent) and, notably, this system assures the
total electroneutrality condition $\sum_i \r_j = 0$. The present approach can be seen as a
more detailed alternative to the phenomenological Thomas-Fermi treatment \cite{smadici}.

It is elementary to resolve Eq. \ref{eq4} for the densities through the doping levels:
$p_j = \sum_{j'} f_{jj'}(\a)x_{j'}$. The problem is resonably simplified considering it
periodic, then the period of $n$ layers at given $\a$ fully defines the coefficients $f_{jj'}
(\a)$ for $1 \leq j,j' \leq n$. For the sake of definiteness, let us consider the sample
system like that in the experiment, Ref. \cite{Boz1}, with $n = 6$ and $x_1 = x_2 = x_5 =
x_6 \equiv x$, $x_3 = x_4 = 0$ (Fig. \ref{fig1}). The explicit solution of Eq. \ref{eq4}
in this case reads:
\bea
 p_1 & = & p_6 = \left(1 - \frac 1{(\a + 1)(\a + 3)}\right)x, \nn\\
  p_2 & = & p_5 = \frac {\a + 2}{\a + 3}x, \nn\\
   p_3 & = & p_4 = \frac {\a + 2}{(\a + 1)(\a + 3)}x,
    \lb{eq6}
     \eea
satisfying the evident electroneutrality condition $p_1 + p_2 + p_3 = 2x$.

\begin{figure}
\center\includegraphics[width=7cm]{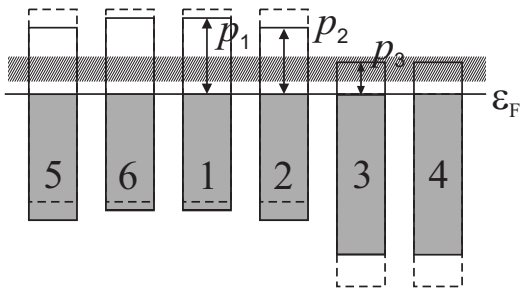}\\
\caption{Modulated electronic configuration by the shifted energy bands (solid rectangles)
in the nanostructured system by Fig. \ref{fig1}, calculated for $x = 0.45$ and localization
parameter $\a = 1$. The dashed rectangles indicate unshifted energy bands for isolated doped
and undoped layers, and the hatched stripe marks the interval of carrier densities where superconductivity should exist.}
  \label{fig2}
\end{figure}

Using the soft X-ray resonant scattering techniques \cite{Goz} for direct measurement of
carrier densities in the experiment, Ref. \cite{smadici}, yielded: $p_1^{\rm exp} \approx 0.33,\,
p_2^{\rm exp} \approx 0.24, \,p_3^{\rm exp} \approx 0.15$. A reasonable fit to this set can be
achieved from Eq. \ref{eq6} with the choice of $\a = 1$: $p_1^{\rm theor} \approx 0.315, \,
p_2^{\rm theor} \approx 0.27, \, p_3^{\rm theor} \approx 0.135$, that is within the experimental
error of $\pm 0.03$ from the measured values.

In order to relate these carrier densities with the experimentally defined critical
temperatures, we can employ the phenomenological bell-like dependence:
\be
 T_{\rm ph}(p) = (p_{\max} - p)(p - p_{\min})T^*,
  \lb{eq7}
   \ee
with $p_{\min} = 0.07, p_{\max} = 0.2$, and $T^* = 9000$ K (this curve being slightly below
of the commonly reported $T_c(p)$ in bulk LSCO \cite{kast}). Using  $p = p_3^{\rm theor}$ in
Eq. \ref{eq7} yields the value of $T_c \approx 38$ K, just that observed in Ref. \cite{smadici}.
This confirms the conclusion that the SC state in this system is limited to the nominally
undoped layers 3,4 as represented schematically in Fig. \ref{fig2}.

One can compare the fitted value of $\a = 1$ with the theoretical expression, Eq. \ref{eq5},
using the standard values $a \approx 0.38$ nm, $c = 1.3$ nm and $W \approx 2$ eV. This
suggests as high value of $\e_{\rm eff}$ as $\sim 150$, however it does not seem unrealistic
if the static \emph{c}-axis polarizability for ${\rm La}_2{\rm CuO}_4$ \cite{chen,reagor} is
enhanced by a contribution from doped mobile carriers.

\begin{figure}
\center\includegraphics[width=9cm]{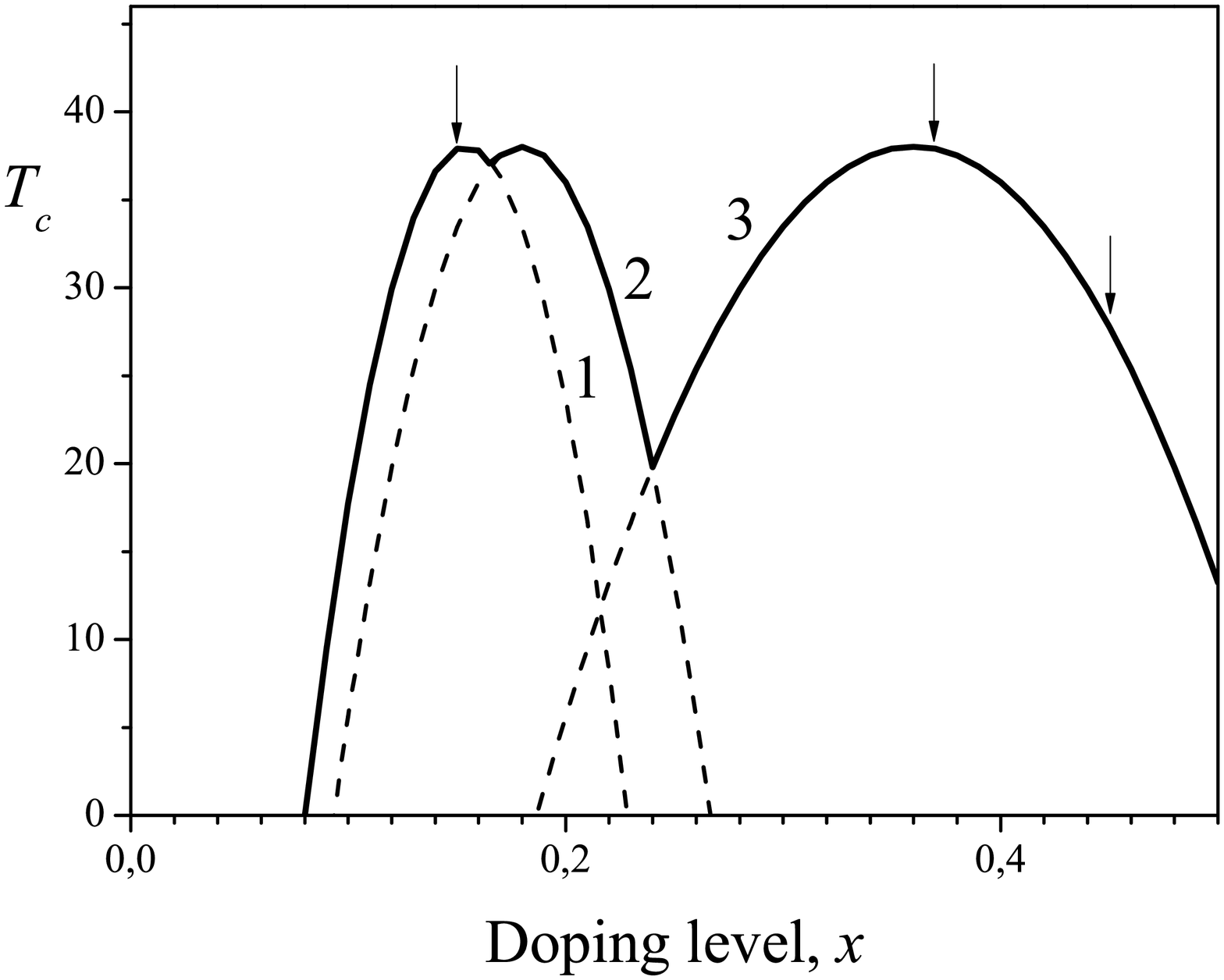}\\
\caption{Calculated critical temperature $T_c$ vs doping level $x$ for the $\left({\rm La}_{2-x}
{\rm Sr}_x {\rm CuO}_4\right)_4({\rm La_2CuO_4})_2$ system. Three numbered curves refer to the 
respective layers from Fig. \ref{fig2} and arrows indicate the doping levels of 0.15, 0.36 
and 0.45 as in the experimental systems \cite{Boz1,smadici}.}
  \label{fig3}
\end{figure}

The situation can be further traced at varying the doping level $x$ (with $\a$ supposedly
constant). Thus, for $x = 0.45$ we obtain respectively: $p_1^{\rm theor} \approx 0.395,\,
p_2^{\rm theor} \approx 0.34, \, p_3^{\rm theor} \approx 0.165$, and then using this
$p_3^{\rm theor}$ in Eq. \ref{eq7} results in $T_c \approx 30$ K, again in agreement with the
measured value \cite{Boz1}.

At least, for the nominally optimum doping level $x = 0.15$, we have: $p_1 = 0.132, \, p_2
= 0.113$ and $p_3 = 0.055$, and the SC state with almost maximum $T_c$ should persist only
in the doped 1,2,5,6 layers separated by the insulating 3,4 layers. This agrees with the
observation of blocked tunneling through the undoped La$_2$CuO$_4$ layer sandwiched between
optimally doped ${\rm La}_{2-x}{\rm Sr}_x{\rm CuO}_4$ electrodes \cite{Boz1}.

Furthermore, combining the results, Eq. \ref{eq6}, and the phenomenological dependence, Eq.
\ref{eq7}, one can easily build a model dependence for critical temperature of SC transition
in the given ${\rm La}_{2-x}{\rm Sr}_x{\rm CuO}_4-{\rm La}_2{\rm CuO}_4$ system vs the doping
level $x$. As seen from Fig. \ref{fig3}, this dependence chosen as the maximum value from
three bell-like curves: $T_c(p) = \max_j T_{\rm ph} (p_j(x))$, has generally a non-monotonous
behavior with the broadest region contributed by the 3,4 layers. It should be noted that the
SC state realized in this region may be of special interest since much longer lifetimes of
charge carriers in the nominally undoped layers, in similarity with the well explored physics
of 2D electron gas (2DEG) inverse layers in semiconducting heterojunctions \cite{2deg}. This
can be an important property for envisaged superconducting devices in nanotaylored heterosystems
\cite{hetero} or excitonic superconductors \cite{ginz}.

\begin{figure}
\center\includegraphics[width=7cm]{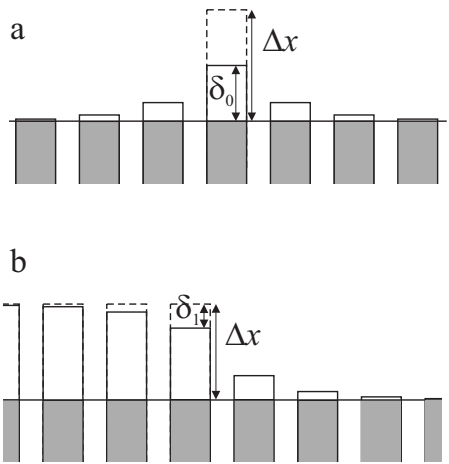}\\
\caption{Modulated electronic configurations for a) single doped layer between undoped
semispaces and b) interface between two semispaces with different doping levels.}
  \label{fig5}
\end{figure}

The model, Eqs. \ref{eq1}-\ref{eq5} can be easily extended to other characteristic
nanostructures. Thus, inclusion into an infinite stack of layers with some uniform doping
level $x$ of a single layer with different level $x + \D x$ will produce a symmetric
distribution of carrier densities $p_j = p_{-j}$ that obviously tend to the asymptotic value:
$p_{j \to\infty} \to x$. Then, considering the reduced densities $\d_j = p_j - x$, we obtain
from Eq. \ref{eq4} an infinite set of linear equations:
\bea
 (2+\a)\d_0 - 2\d_1 & = & \a \D x,\nn\\
  (2+\a)\d_j = \d_{j+1} & + & \d_{j-1},\qquad j\geq 1,
   \lb{eq8}
    \eea
with the electroneutrality condition $\d_0 + 2\sum_{j\geq 1}\d_j = \D x$. It can be easily
checked that the system, Eq. \ref{eq8}, is solved with $\d_j = \d_0 \exp(-\kappa j)$ where
$\kappa = {\rm arccosh}(1+\a/2)$ and with the most interesting central value given by $\d_0 =
\D x \tanh \kappa/2 = \D x\sqrt{\a/(a+4)}$. From this function, it follows that the greatest
part of added charge density remains at the central layer, $\d_0 > \D x/2$, when the localization
parameter $\a$ surpasses $4/3$. Though being somewhat higher of that used in the previous
analysis of periodically doped system, such value can be supposed to describe a stronger
localization for the single layer doping. Then it can support the experimental observation
of persisting SC state in a single optimally doped layer sandwiched between undoped
semispaces ($x = 0,\,\D x = x_{\rm opt}$) \cite{Boz4}, if $p_0$ falls within the range
$\left[p_{\min},p_{\max}\right]$. Contrarywise, a single undoped layer ($x_0 = 0$) between
optimally doped semispaces ($x = x_{\rm opt} = - \D x$) should possess a lower local density
$p_0 = x_{\rm opt} - \d_0$, which more probably goes out of $\left[p_{\min},p_{\max}\right]$
so that this layer would pertain insulating.

Another exemplary case is the interface between two semi-infinite stacks of layers with
different uniform doping levels $x_j = x$ at $j \leq -1$ and $x_j = x - \D x$ at $j \geq
1$, where the reduced densities can be defined respectively as $\d_j = x_1 - p_j$ at $j \leq -1$
and $\d_j = p_j - x_2$ at $j \geq 1$ with evident symmetry $\d_j = \d_{-j}$. Then Eq. \ref{eq8}
is reformulated as:
\bea
 (3 + \a)\d_1 - \d_2 & = & \D x = x_1 - x_2,\nn\\
  (2+\a)\d_j = \d_{j+1} & + & \d_{j-1},\qquad j\geq 2,
   \lb{eq9}
    \eea
and its solution is given by $\d_j = \d_1 \exp[-\kappa (j-1)]$ with the same $\kappa$ as above
and with $\d_1 = \D x/\left[2 - \a/2 + \sqrt{\a(\a + 4)}\right]$.

At least, combining the previous cases can serve to explain the "giant proximity effect"
observed in a thick underdoped layer sandwiched between optimally doped electrodes \cite{Boz5}.

In conclusion, a simple electrostatic model model combined with 2D electronic band spectrum
is used to semi-quantitatively explain the recent experimental findings in ${\rm La}_{2-x}
{\rm Sr}_x{\rm CuO}_4$ multilayered systems, selectively doped with precision to single
atomic layer. Exact solutions are found for local charge densities $p$ in conducting CuO$_2$
planes, for a number of periodic and non-periodic doping configurations, permitting agreement
with the experimentally defined $p$'s and SC transition critical temperature $T_c$. The model
can be used for effective designing of new SC (including Josephson tunnel) systems.

\end{document}